\documentclass[12pt]{article}

\usepackage[
  a4paper, mag=1000,
  left=2cm, right=2cm, top=2cm, bottom=2cm, headsep=0.7cm, footskip=1cm
]{geometry}

\usepackage{indentfirst}

\usepackage[T2A]{fontenc}
\usepackage[utf8]{inputenc}
\usepackage[english]{babel}
\usepackage{amsmath,amssymb,amsfonts}
\usepackage{graphicx}
\usepackage{epstopdf}
\usepackage[autostyle]{csquotes}
\usepackage[noblocks]{authblk}
\usepackage{xcolor}
\definecolor{darkblue}{rgb}{0,0,0.5}
\usepackage[colorlinks,linkcolor=darkblue]{hyperref}
\usepackage{subcaption}
\usepackage{cite}

\newcommand{\bra}[1]{\left\langle#1\right|}
\newcommand{\ket}[1]{\left|#1\right\rangle}

\newcommand{\mean}[1]{\left\langle {#1}  \right\rangle}
\newcommand{\cs}[1]{$\mathrm{#1}$}
\newcommand{\ua}{\uparrow}
\newcommand{\da}{\downarrow}

\begin{document}

\title{Localization Transition on Random Graphs with Chiral and Bogoliubov-de Gennes Symmetry Classes}
\author[1,2]{Daniil Kochergin}
\affil[1]{Phystech School of Applied Mathematics and Computer Science, \newline Moscow Institute of Physics and Technology, Institutsky lane 9, \newline Dolgoprudny, Moscow region, 141700}
\affil[2]{Laboratory of Complex Networks, Center for Neurophysics and\newline Neuromorphic Technologies, Moscow, Russia}

\date{\today}

\maketitle

\begin{abstract}
    We studied single-particle Anderson localization in ensembles of graphs that correspond to chiral and Bogoliubov-de Gennes (BdG) symmetry classes. For a random biregular bipartite graph with chiral symmetry, the density of states was found using the cavity approach. Calculating the fractal dimension shows the effects of disordered zero modes. For Bogoliubov-de Gennes ensembles with an underlying random regular graph (RRG), the density of states was calculated both numerically and analytically. The ensembles BdG-RRG with symmetry-conserving diagonal disorder in the delocalized phase have a smaller fractal dimension compared to the usual RRG.
\end{abstract}

\section{Introduction}

Symmetry classification provides a division of all Hermitian disordered systems into groups composed of three Wigner-Dyson classes, three chiral ensembles, and four Bogoliubov-de Gennes (BdG) ensembles, which differ in the presence of time-reversal and spin rotation or combinations of them~\cite{Altland1997Nonstandard}. The ten-fold classification allows for a systematic approach to studying single-particle Anderson localization based on the underlying Hamiltonian symmetry~\cite{anderson1958,Mirlin2010Anderson}. Furthermore, important effects for system properties play not only the symmetry class but also topological invariants~\cite{Qi2011Topological}.

Localization in finite-dimensional periodic systems or random matrices has been studied in a variety of studies~\cite{Wang2021Chiral,Culver2025Universal,Geier2020Symmetry,Fan2020Superconductivity,Beenakker2015Random}. In the article \cite{GarciaGarcia2006Anderson}, for three-dimensional disordered systems, the  chiral symmetry induced Anderson localization with a mobility edge near zero affects critical properties by additional symmetry. The phase diagram of the topological properties of the localized and delocalized phases for all ten symmetry classes for noninteracting fermionic quasiparticles was established~\cite{Morimoto2015Anderson}. 
The superconducting Bardeen-Cooper-Schrieffer Hamiltonian in the mean-field approximation has BdG symmetry classes. 
In the article~\cite{Burmistrov2012Enhancement} for the two-dimensional system in weak localization and weak antilocalization regimes, it was shown that Anderson localization leads to an increase in the critical temperature for the superconducting transition. 

Anderson localization has notable applications in many-body interaction systems, where in the case of long-range interaction in zero-approximation many-body localization can be seen as spinless single-particle localization with on-site disorder on a Bethe lattice or Random Random Graph (RRG), with vertices associated with many-body states and edges to transitions between states \cite{altshuler1997quasiparticle, Tikhonov2021From}. 
The localization transition on the graph matches the transition from ergodicity to many-body localization. In the context of symmetry classes, most studies of localization on graphs consider graphs that are discrete versions of the Gaussian Orthogonal Ensemble (GOE), with \cs{AI} symmetry class with time-symmetry and without spin-symmetry.
The localization properties of the diagonal disorder-free case for random bipartite (chiral) graphs were studied in~\cite{MartínezMartínez2019}, the graphs with power-law degree distribution in~\cite{Slanina2017Localization}, and the investigation of the spectral gap for biregular random graphs in~\cite{Brito2022Spectral}.
However, studying other symmetry classes in the context of many-body localization is limited to many-body Hamiltonian~\cite{Finelli2002Nuclear,zhang2025zero}. In that report, we study the localization transition of graphs for random (bi-)regular graphs with chiral and BdG symmetry classes combined with Anderson's on-site disorder. 

In the article, the ensembles of random Hamiltonians $H$ are studied.
For each Hamiltonian, the Schrödinger equation has the form
\begin{equation}
    H \psi_i = \lambda_i \psi_i \ ,
\end{equation}
where $\lambda_i$ are the eigenvalues of the $\psi_i$ eigenstates.
From a spectral point of view, we will study the level spacing distribution (LSD) $P(s)$ between nearest energy levels,
\begin{equation}\label{eq:LS}
 s_i=E^{(u)}_{i+1}-E^{(u)}_i \ ,
\end{equation}
where $E^{(u)}_{i}$ are energies after spectrum unfolding. For unfolding, we find the cumulative distribution function ($\mathrm{CDF}$) as the fraction of eigenvalues that are less than $\lambda$, ensemble average of $n^{real}$ samples of random Hamiltonians of size $N\times N$, $\lambda_i^{(j)}$, $1\leq j\leq n^{real}$:
\begin{equation}
\mathrm{CDF}(\lambda)=\frac{\#(\lambda_i^{(j)}<\lambda)}{n^{real}N} \ .
\end{equation}
As a result,
\begin{equation}\label{eq:E_i_unfolded}
 E^{(u)}_i=\mathrm{CDF}(\lambda_i)N \ .
\end{equation}
LSD forms the Wigner surmise for ergodic states and has Poisson statistics for localized states~\cite{Shklovskii1993Statistics}.

From the eigenvector perspective, we consider the inverse participation ratio (IPR), which for $q=2$ describes the fraction of nodes that participate in the state, 
\begin{equation}
    IPR_q=\sum_n^N |\psi_{\lambda}(n)|^{2q}  \ .
\end{equation}
and the corresponding fractal dimension that can be calculated with $IPR_q$ as
\begin{equation}
    D_q=-\frac{\log_2 IPR_q}{(q-1)\log_2 N} \ .
\end{equation}
Fractal dimension for localized states $D_q \to 0$ and for ergodic states $D_q \to 1$~\cite{Schreiber1991Multifractal}. Intermediate values, which correspond to nonergodic delocalized states, are out of the scope of the current discussion.

\section{Chiral ensemble}\label{Sec:Chiral}
Since the point of our interest is weightless undirected random graphs, for chiral classes, we consider the \cs{BDI} class that corresponds to the chiral orthogonal ensemble with time-reversal and spin rotation symmetries. The graph that represents the \cs{BDI} class consists of two sets of nodes with sizes $N$ and $M$ and has edges only between the vertices in different sets. Edges between nodes in the same set are forbidden, i.e. it is a bipartite graph.
The Hamiltonian in the diagonal disorder-free case, or in other words, the adjacency matrix, can be represented as follows.
\begin{equation}\label{eq:chAdj}
    A=
    \begin{pmatrix}
        0 & t \\
        t^T & 0 \\
    \end{pmatrix}\ , 
\end{equation}
where $t$ is a matrix of size $N\times M$, consisting of~$0$ and~$1$, which are distributed randomly according to ensemble properties. It obeys the equation
$\tau_z H \tau_z = - H$, where $\tau_z$ is the third isospin Pauli matrix. Since for Fock space, the usual approximation is graphs that have a constant degree distribution , i.e. RRG, we choose an ensemble of random biregular bipartite graphs, with $\sum_m t_{nm} = d_N$ and $\sum_n t_{nm} = d_M$, $d_N N = d_M M$, where $d_N$ and $d_M$ are partition node degrees. For simplicity, let us call them chiral RRG.

\begin{figure*}[ht]
    \centering
    \includegraphics[width=0.33\linewidth]{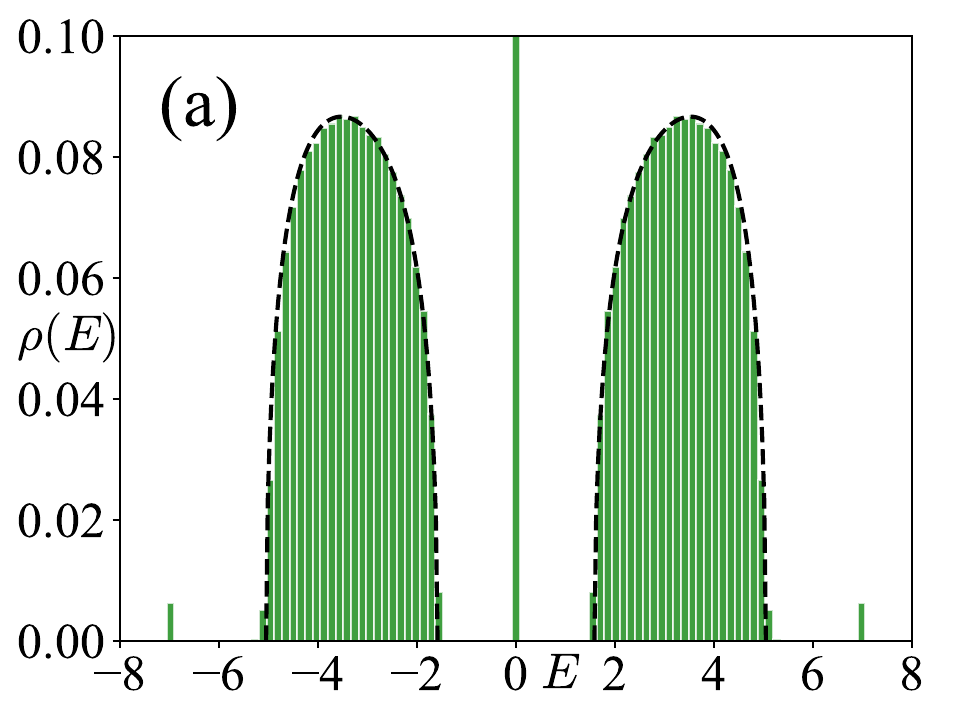}%
    \includegraphics[width=0.33\linewidth]{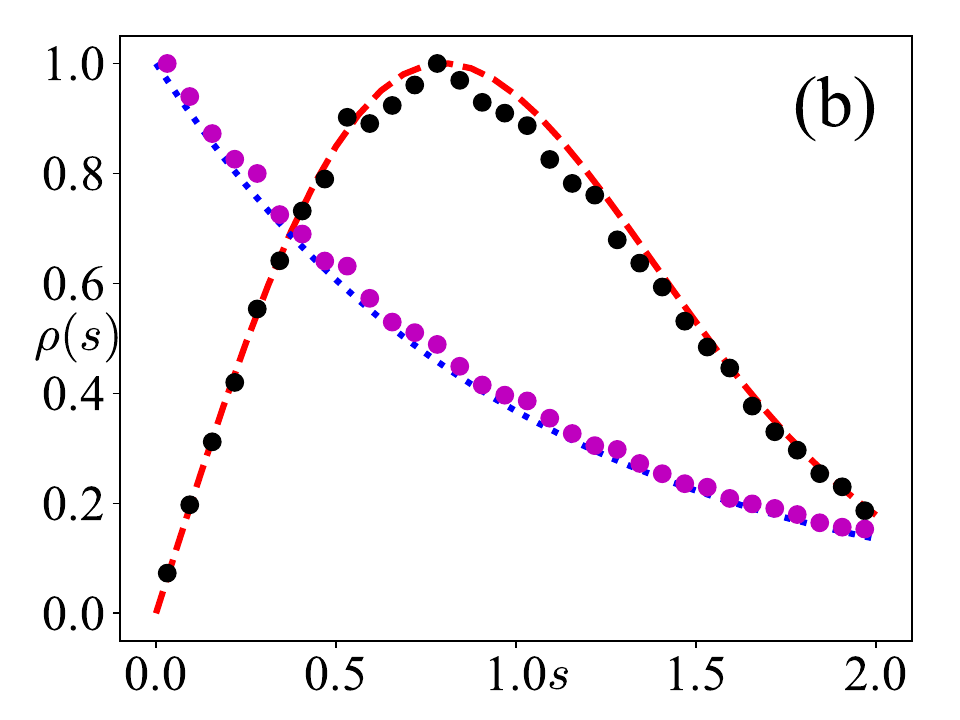}%
    \includegraphics[width=0.33\linewidth]{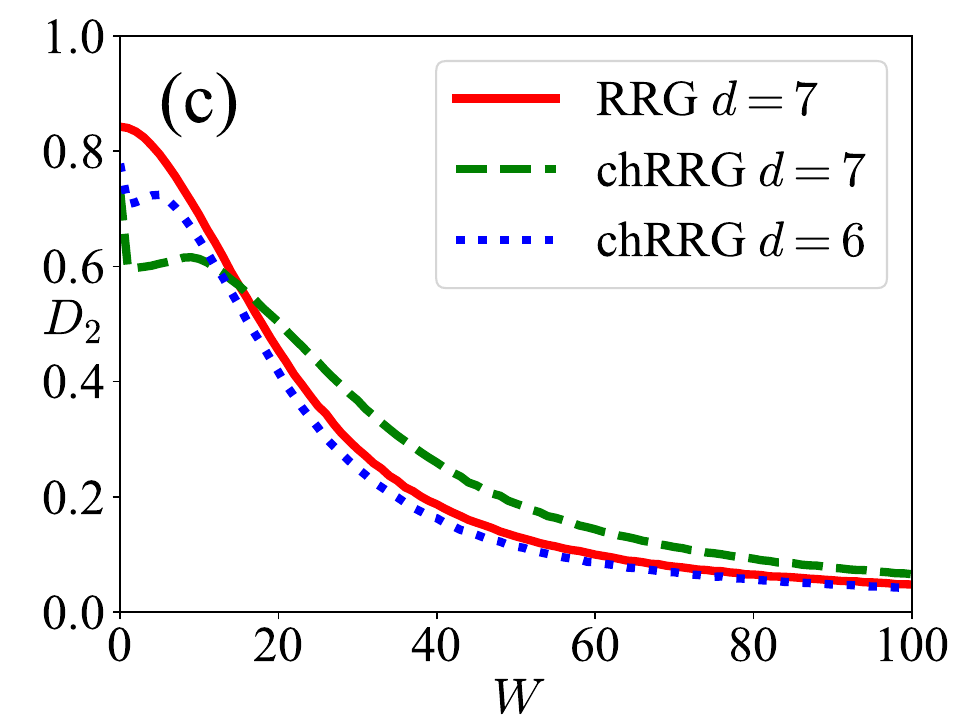}
    \includegraphics[width=0.33\linewidth]{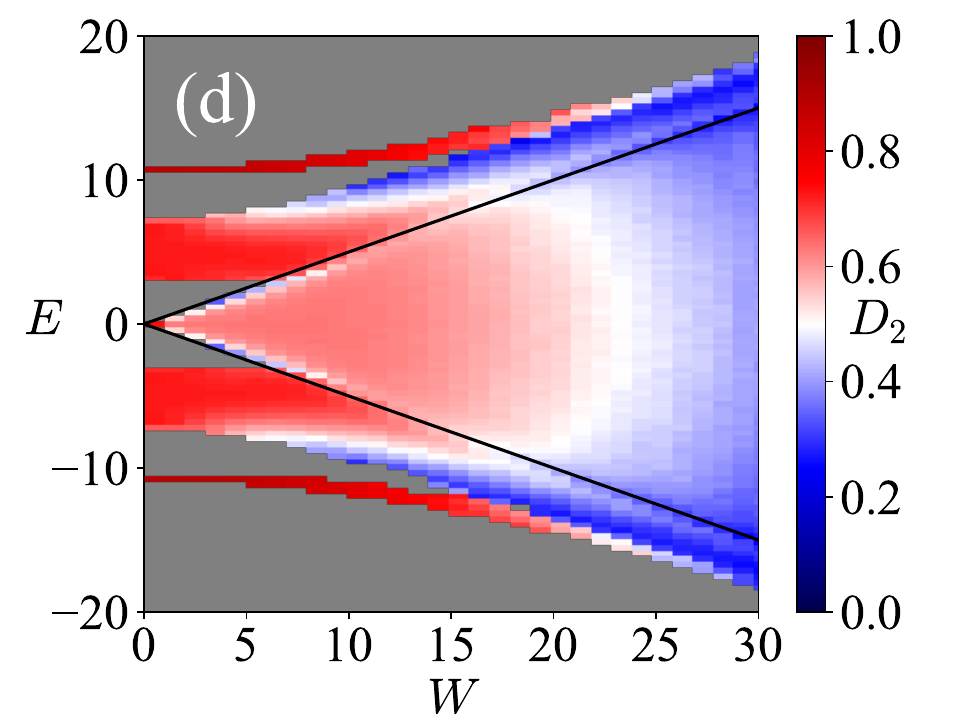}%
    \includegraphics[width=0.33\linewidth]{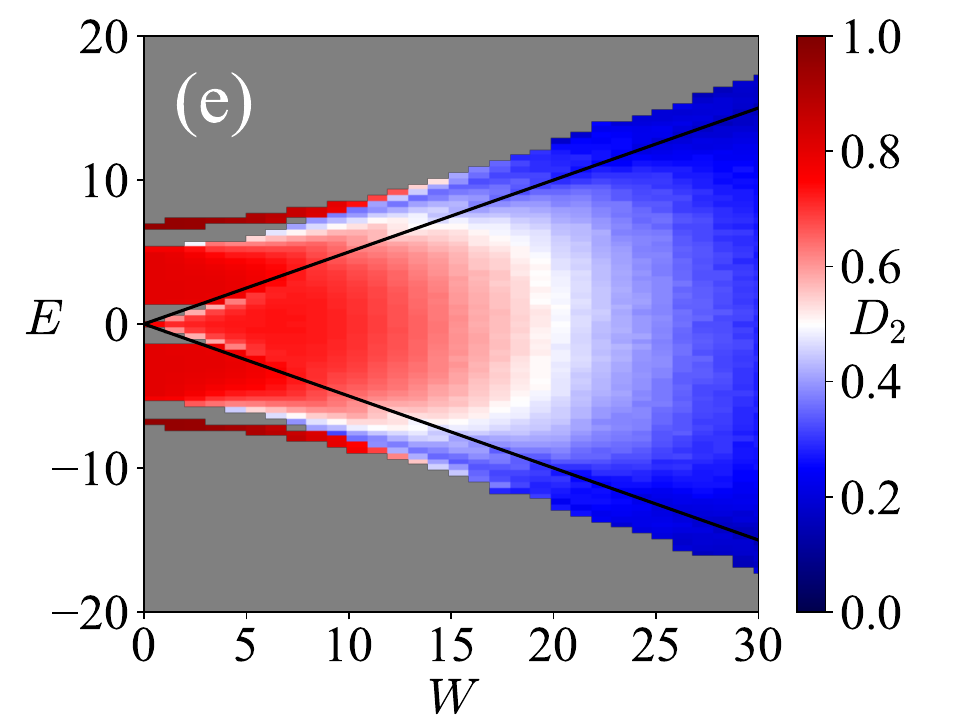}%
    \includegraphics[width=0.33\linewidth]{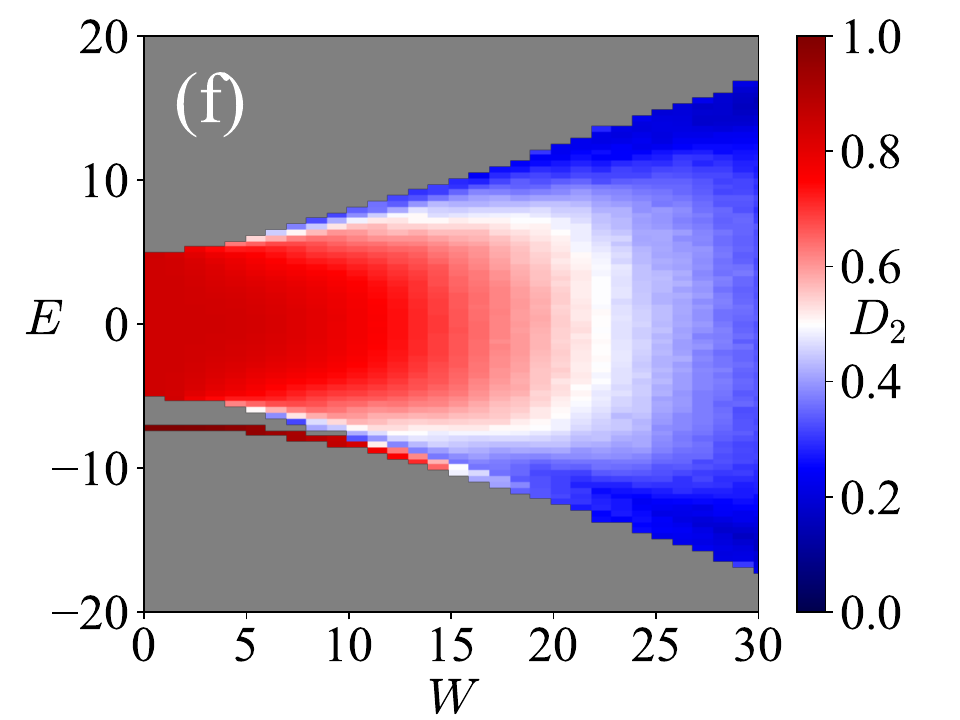}
    \caption{Numerical calculation of (a) density of states and (b) level spacing distribution for bipartite biregular graph with partition $N = 256$, $M = 768$, $d_N = 12$, $d_M = 4$.
    The black dashed line shows DOS calculated using Eq.~(\ref{eq:DOS_ch}). The purple and black dots indicate LSD at localized ($W = 1000$) and delocalized regimes. Blue and red lines show Poisson and Wigner-Dyson distribution. (c)~Comparison of average fractal dimensions $D_2$ for different ensembles depending on diagonal disorder $W$. Panels (d)-(e) shows dependence of fractal dimension $D_2$ on energy $E$ and diagonal disorder $W$.}
    \label{fig:1}
\end{figure*}

Since we want study the level spacing first consider the density of states for the system. The cavity approach~\cite{abou1973selfconsistent} allows to obtain the system of equations on the single-site Green functions $G^{N(M)}$, where the subscript $N$ or $M$ represents the partition with $N$ or $M$ nodes.
\begin{equation}
    \begin{cases}
        \frac{1}{G^N_n} = E + i \eta - \sum_m^{d_N} |t_{nm}|^2 G^{M}_{m\to n} \\
        \frac{1}{G^M_m} = E + i \eta - \sum_n^{d_M} |t_{nm}|^2 G^{N}_{n\to m} \\
    \end{cases} \ ,
\end{equation}
where $E$ is the energy and $\eta \to 0$. $G^{M}_{m\to n}$ and $G^{N}_{n\to m}$ are Green tree functions for the graph without the edge from $n$ to $m$. For an ensemble of random biregular bipartite graphs, we can use average Green functions $G^{N(M)} = \mean{G^{N(M)}_k}$ and $G^{N(M)}_{\to} = \mean{G^{N(M)}_{k\to l}}$. As a result:
\begin{equation}
    \begin{cases}
        \frac{1}{G^N} = E + i \eta - d_N G^{M}_{m\to n} \\
        \frac{1}{G^M} = E + i \eta - d_M G^{N}_{n\to m} \\
    \end{cases} \ .
\end{equation}
Since RRG has a local tree-like structure, $G^{N(M)}_{\to}$ can be found using self-consistent equations.
\begin{equation}
    \begin{cases}
        \frac{1}{G^N_{\to}} = E + i \eta - k_N G^{M}_{\to} \\
        \frac{1}{G^M_{\to}} = E + i \eta - k_M G^{N}_{\to} \\
    \end{cases} \ .
\end{equation}
where branching $k_N=d_N-1$.
For the tree Green's function the solution is
\begin{equation}
    G^N_{\to} = \frac{(E^2+k_M-k_N) \pm \sqrt{(k_N - k_M - E^2)^2-4 k_M E^2}}{2k_M E}
\end{equation}
and $G^M_{\to}$ can be found similarly. 
For a random bipartite biregular graph, it is now possible to find $G^{N}$ and
\begin{equation}
    G^M = \frac{ 2 k_M E^2 + d_M (k_N - k_M - E^2)}{2 E (d_M(k_N-k_M) + d_N^2 - E^2)} 
    \pm
    \frac{d_M \sqrt{(k_N-k_M-E^2)^2-4 k_M E^2}}{2 E (d_M(k_N-k_M) + d_N^2 - E^2)} \ .
\end{equation}
Finally, the density of states becomes
\begin{equation}\label{eq:DOS_ch}
    \rho(E) = \frac{1}{\pi}\operatorname{Im} \mean{G} = 
            \frac{N \operatorname{Im} G^N + M \operatorname{Im}  G^M}{\pi(N+M)}\ .
\end{equation}
The resulting formula coincides with the density of states in the article~\cite{Godsil1988Walk}. Except for two continuous spectrum parts, described by the Eq.~(\ref{eq:DOS_ch}), there are $|N-M|$ zero modes, and the Perron eigenvalue and its symmetrical counterpart s with energies $E = \pm\sqrt{d_Nd_M}$.

Figure~\ref{fig:1}(a) shows numerical calculations for the density of states. From the eigenvalue point of view, the LSD has identical localization properties to RRG (Fig.~\ref{fig:1}(b)), Wigner-Dyson distribution for the delocalized phase, and Poisson for the localized.

Figure~\ref{fig:1}(c) compares the average fractal dimensions for different systems; for these ensembles, the dependencies of the fractal dimension on energy and disorder's amplitude are presented on~Fig.~\ref{fig:1}(d)-(e). 
The first is chiral RRG with $N = 256$, $M = 768$, $d_N = 12$, $d_M = 4$, it has mean degree $d = 6$ (blue line, panel~(e)). 
The second is RRG with $N+M$ nodes and $d=6$ (black line, panel~(e)).
The last is again a chiral RRG with $N = 128$, $M = 896$, $d_N = 28$, $d_M = 4$, $d = 7$ (green line). 
The last two have approximately the same maximal energy as that of the continuous spectrum part. $E^{\text{chRRG}}_{\text{max}} = \sqrt{d_N-1} + \sqrt{d_M-1} \approx 5.05$ and $E^{\text{RRG}}_{\text{max}} = \sqrt{4 (d-1)} \approx 4.90$. 

For the chiral RRG with diagonal disorder with uniform distribution over the range $(H_{ch+A})_{nn} = \varepsilon_n \in [-W/2, W/2]$, the Anderson transition does not drastically differ from that in the RRG case. The main difference lies in the presence of zero modes that appear when $|N-M|>0$. They are unstable to the diagonal disorder at the range $E=[-W/2,W/2]$, indicated by the black solid line on panels (d) and (e). When $N-M$ is near a higher value, disordered zero modes become dominant in the average IPR compared to states from the continuous spectrum part. That leads to a smoother transition for the chiral RRG (the green dashed line) compared to the RRG (the red solid line) with equal mean node degrees. For systems with approximately the same maximal energy of central bands, the fractal dimension is slightly smaller for the chiral graph (the blue dotted line).

\section{Superconducting ensemble}\label{Sec:Super}

For superconducting ensembles, let us consider the tight-binding Hamiltonian of spin-1/2 electrons with two-body interactions in the second quantized form:
\begin{equation}
    H = H_0 + H_I = \sum_{\bar i} \epsilon_{\bar i} c^\dagger_{\bar i}c_{\bar i} + 
        \sum_{\bar i \bar j  \bar k \bar l} V_{\bar i \bar j  \bar k \bar l} 
                c^\dagger_{\bar i} c^\dagger_{\bar j} c_{\bar k} c_{\bar l} \ ,
\end{equation}
where indices with overline, $\bar i$, are combinations of node and spin. In the Hamiltonian, the first term is on-site disorder, which in general can be different for opposite spin directions. The second term is the interaction of electron pairs with all possible spins and sites. 

The general way to construct a single-particle Hamiltonian for a system with many-body interactions is to change basis to Fock states, where eigenstates will have the following form:
\begin{equation}\label{eq:FS_p}
    \ket{\alpha} = \ket{\alpha_\ua} \ket{\alpha_\da} = \prod_k^L (c^\dagger_{k,\uparrow})^{n_{k,\uparrow}}
                   \prod_k^L (c^\dagger_{k,\downarrow})^{n_{k,\downarrow}} \ket{0} \ ,
\end{equation}
where $n_{k,\sigma}$ is the number of electrons at site $k$ with spin $\sigma$, that is, it can have values $0$ or $1$. $\ket{0}$ is the vacuum state. $L$ is the size of the system. The total number of states and, consequently, the number of graph nodes is $N = 2^{2L}$.

From the interaction term of the Hamiltonian, the possible configurations for spin-1/2 electron pairs are $\ua\ua$, $\da\da$, $\ua\da$, and $\da\ua$. 
First, we consider terms where the electron pairs have the same spin in all creation and annihilation operators. If the interaction has spin-up, then it does not change the spin-down sectors in the configuration of the Fock state and vice versa. 
\begin{equation}
\begin{gathered}
     A_{\alpha\beta} = 
        \bra{\alpha_\ua} \sum_{i j k l} V_{\bar i \bar j  \bar k \bar l}  c^\dagger_{i\ua} c^\dagger_{j\ua} c_{k\ua} c_{l\ua} \ket{\beta_\ua} \ , \\
     C_{\alpha\beta} =  
        \bra{\alpha_\da} \sum_{i j k l} V_{\bar i \bar j  \bar k \bar l}  c^\dagger_{i\da} c^\dagger_{j\da} c_{k\da} c_{l\da} \ket{\beta_\da} \ , \\
        E_{\alpha\ua} =  A_{\alpha\alpha} + \sum_k n_{k\ua} \epsilon_{k\ua} \ ,  \quad 
        E_{\alpha\da} =  C_{\alpha\alpha} + \sum_k n_{k\da} \epsilon_{k\da} \ .
\end{gathered}
\end{equation}
Second, if the interaction term consists of the annihilation of the pair of electrons with spin-down (up) but the pair of electrons with spin-up (down) is created, then both spin sectors of the Fock state are changed.
\begin{equation}
    B_{\alpha\beta} = \bra{\alpha_\da} \bra{\alpha_\ua} \sum_{i j k l} V_{\bar i \bar j  \bar k \bar l}  c^\dagger_{i\ua} c^\dagger_{j\ua} c_{k\da} c_{l\da} \ket{\beta_\ua}\ket{\beta_\da}
\end{equation}
If the Hamiltonian has only the above terms, then in the single-particle matrix representation it looks as follows.
\begin{multline}
    H =  \sum_{\bar \alpha} E_{\bar \alpha} c^\dagger_{\bar \alpha} c_{\bar \beta} + \sum_{\alpha\beta} 
        \left( A_{\alpha\beta} c^\dagger_{\alpha\ua} c_{\beta\ua} + C_{\alpha\beta}  c^\dagger_{\alpha\da} c_{\beta\da}\right) 
      + \sum_{\alpha\beta}  \left(B_{\alpha\beta} c^\dagger_{\alpha\ua} c_{\beta\da} +
                                  B^T_{\alpha\beta} c^\dagger_{\alpha\da} c_{\beta\ua}\right) \ ,
\end{multline}
\begin{equation}
    H = 
    \begin{pmatrix}
        c^\dagger_{\ua} & c^\dagger_{\da} 
    \end{pmatrix}
    \begin{pmatrix}
        A & B \\
        B^T & C \\
    \end{pmatrix}
    \begin{pmatrix}
        c_{\ua} \\ c_{\da} 
    \end{pmatrix}\ .
\end{equation}

The remaining interaction terms consist of the electron pairs, at least one of which has a configuration $\ua\da$ or $\da\ua$. These terms can be simplified using the usual mean-field approximation for the Bardin-Copper-Shriffer theory. We assume that 
\begin{math}
    \mean{c^\dagger_{i\sigma}c^\dagger_{j\sigma}} = \mean{c_{i\sigma}c_{j\sigma}} = 0
\end{math}
and only
$ \mean{c^\dagger_{i\sigma}c^\dagger_{j\sigma'}} $ and $ \mean{c_{i\sigma'}c_{j\sigma}} $ can be different from zero.
The significant part of the interaction will contain the following terms.
\begin{equation}
\begin{gathered}
    \sum_{i j} \Delta_{ij} ( c_{i\downarrow} c_{j\uparrow} + c_{i\downarrow} c_{j\downarrow} ) + 
    \sum_{i j} \Delta^*_{ij} ( c^\dagger_{i\uparrow} c^\dagger_{j\downarrow}  + 
                               c^\dagger_{i\downarrow} c^\dagger_{j\downarrow} )  \ , \\ \quad
    \Delta_{kl} =\sum_{ij} V_{ijkl}  \mean{c^\dagger_{i\uparrow}c^\dagger_{j\downarrow}} \ , \quad
    \Delta^*_{ij} = \sum_{kl} V_{ijkl} \mean{c_{k\downarrow}c_{l\uparrow}} \ \ .
\end{gathered}
\end{equation}

The standard approach to single-particle BdG ensembles introduces a redundant hole counterpart to treat the gap term as noninteracting. To reflect this, we also add the hole part to the state of (\ref{eq:FS_p}).
\begin{equation}
        \ket{\alpha^\text{ph}} = 
        \ket{\alpha^\text{p}_\ua} \ket{\alpha^\text{p}_\da} \ket{\alpha^\text{h}_\ua} \ket{\alpha^\text{h}_\da} =
        \prod_k^L (c^\dagger_{k,\ua})^{n_{k,\ua}} \prod_k^L (c^\dagger_{k,\da})^{n_{k,\da}} \ket{0}
        \prod_k^L (c_{k,\ua})^{n_{k,\ua}}\prod_k^L (c_{k,\da})^{n_{k,\da}} \ket{1} \ ,
\end{equation}
where $\ket{1}$ is the state with the maximal number of electrons. Then the gap matrix on the Fock basis has elements.
\begin{equation}
\begin{gathered}
    \Delta_{\alpha\beta} 
    = \bra{\alpha^{\text{p}}_\ua} \bra{\alpha^{\text{h}}_\da} 
        \sum_{ij} \Delta_{ij} c_{i\da} c_{j\ua} 
      \ket{\beta^{\text{h}}_\da}  \ket{\beta^{\text{p}}_\ua}
    \ , \\
    \Delta^{\ua\ua}_{\alpha\beta} 
    = \bra{\alpha^{\text{p}}_\ua} \bra{\alpha^{\text{h}}_\ua} 
        \sum_{ij} \Delta_{ij} c_{i\ua} c_{j\ua} 
      \ket{\beta^{\text{h}}_\ua}  \ket{\beta^{\text{p}}_\ua} \ .
\end{gathered}
\end{equation}
For the element $\Delta_{\alpha\beta}$, the sum in the definition contains a set of nonzero elements $\Delta_{ij}$; for element $\Delta_{\beta\alpha}$, the sum must have the set of transposed elements $\Delta_{ji}$, which due to fermionic statistics is asymmetric, $\Delta_{ji} = - \Delta_{ji}$. Hence, the gap matrix in the Fock basis inherits the asymmetry $\Delta_{\alpha\beta} = - \Delta_{\beta\alpha}$.

\begin{figure*}[t]
    \centering
    \includegraphics[width=0.33\linewidth]{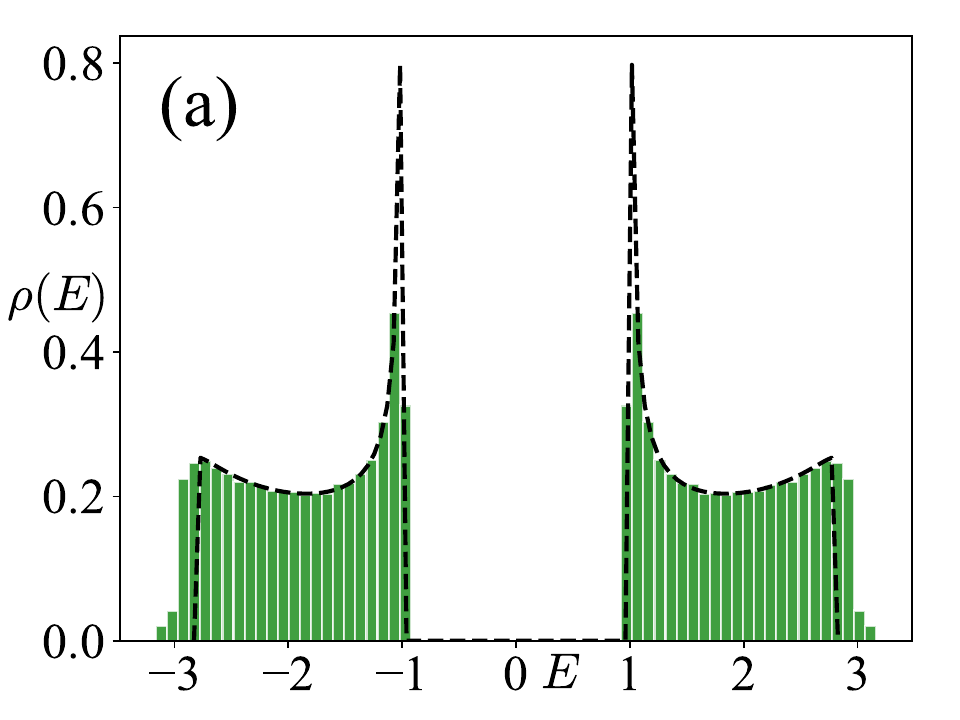}%
    \includegraphics[width=0.33\linewidth]{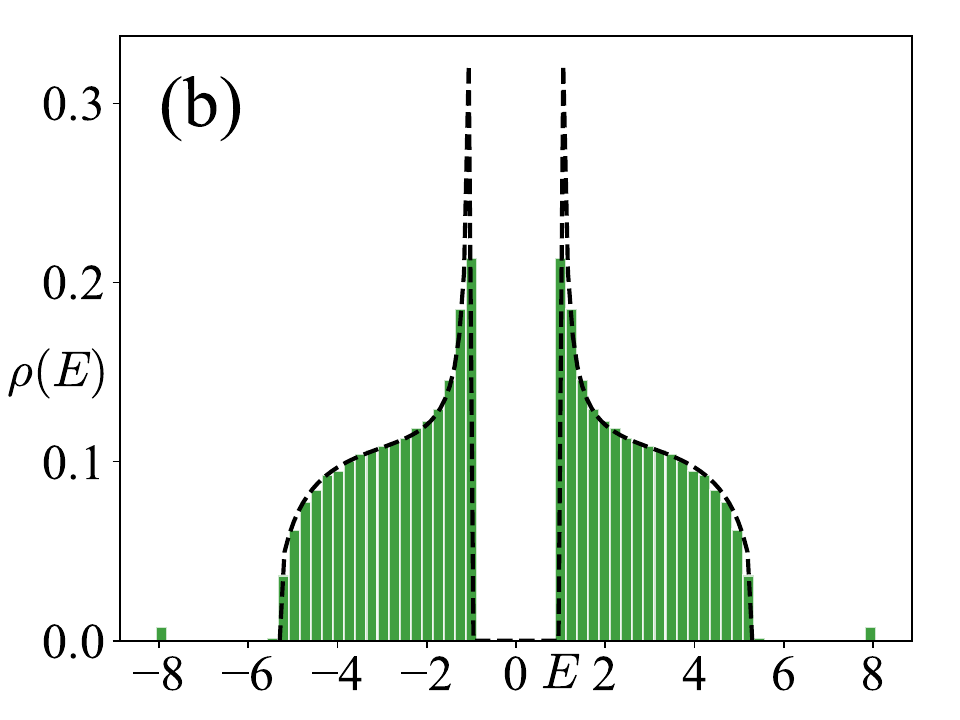}%
    \includegraphics[width=0.33\linewidth]{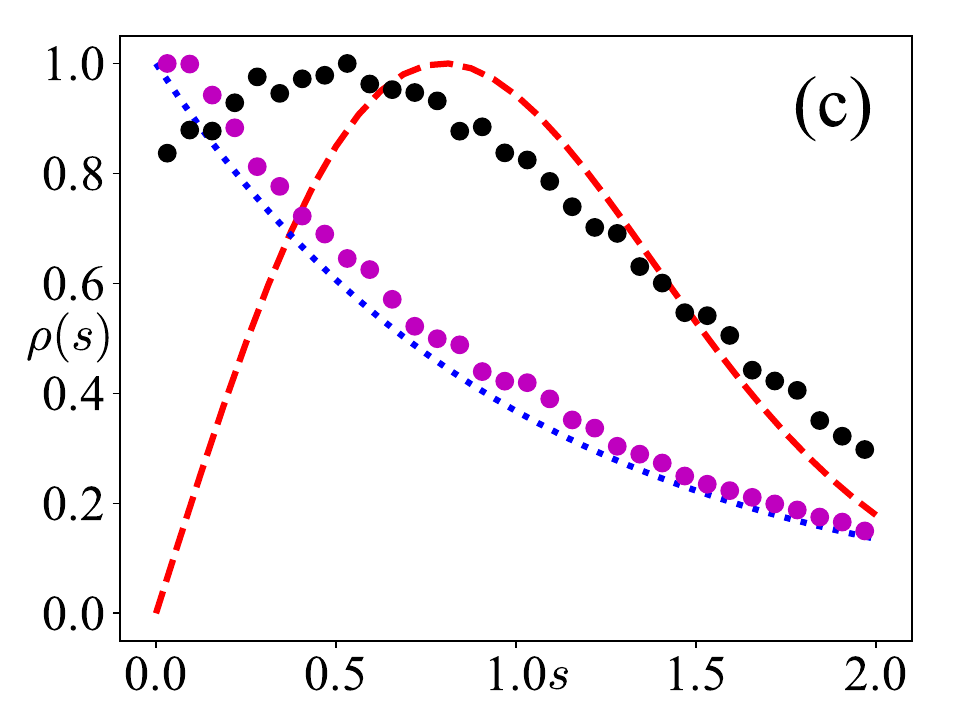}
    \vfill
    \includegraphics[width=0.33\linewidth]{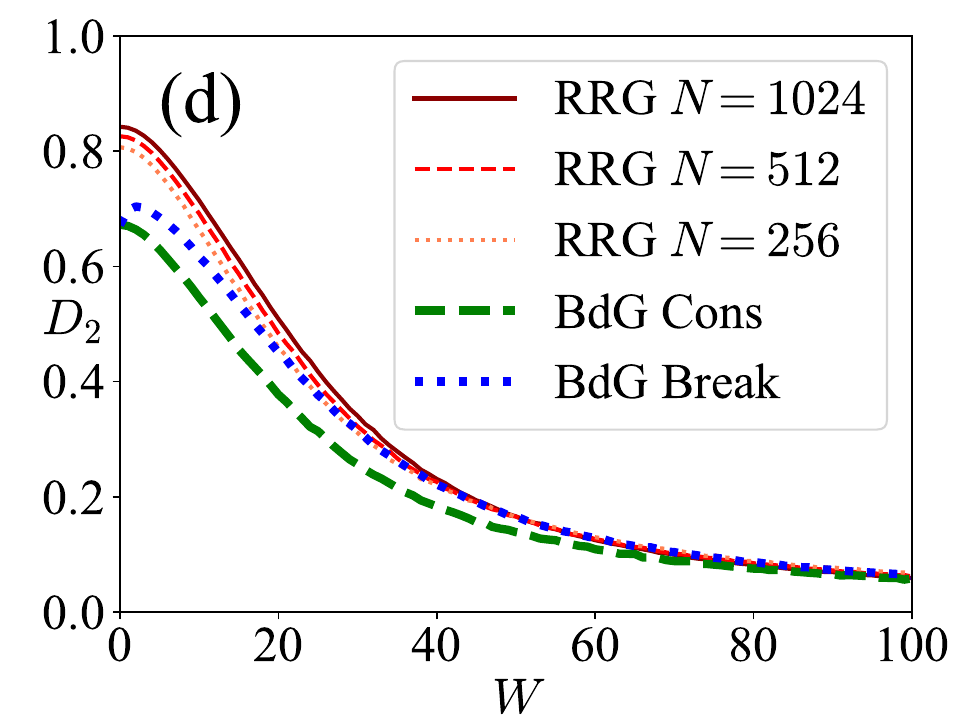}%
    \includegraphics[width=0.33\linewidth]{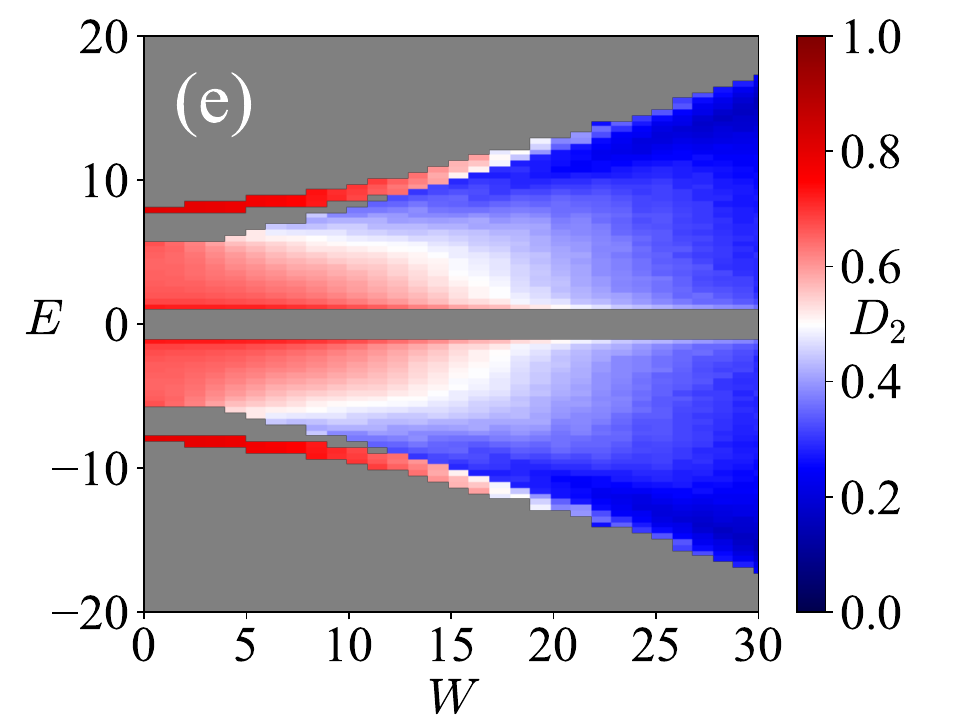}%
    \includegraphics[width=0.33\linewidth]{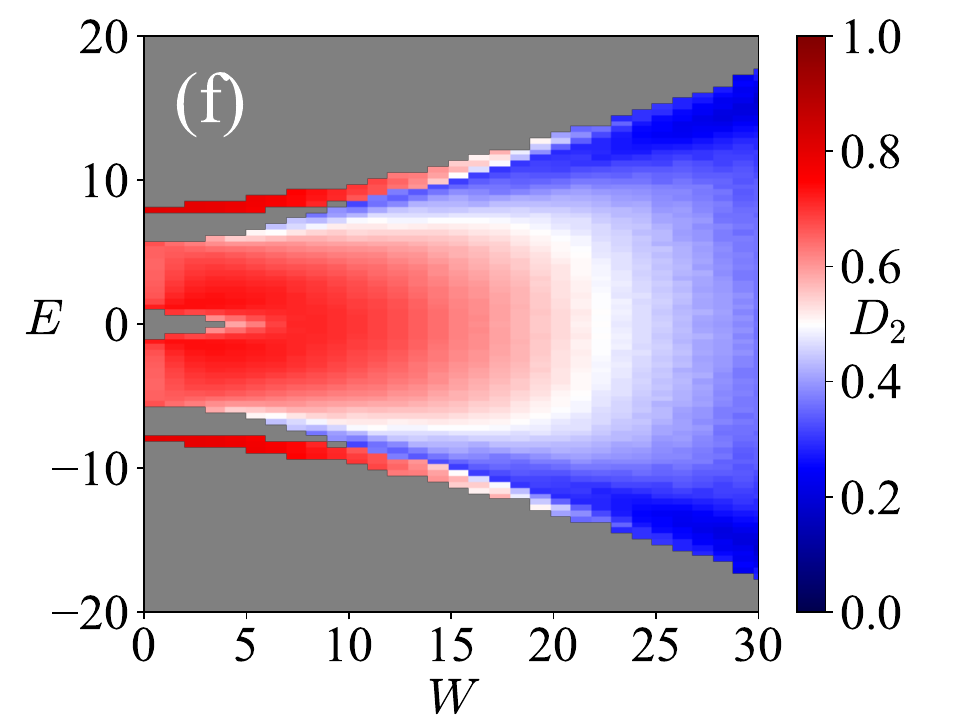}
    \caption{The green histograms shows numerical calculation of density of states for Hamiltonian described by Eq.~(\ref{eq:BdG+A}) without on-site disorder $\Delta_0 = 1$ (a)~$d=3$, (b)~$d=8$, the black dashed lines show DOS calculated by Eq.~(\ref{eq:BdG-KM}). (c)~Level spacing distribution, the purple and black dots indicate LSD at localized ($W = 1000$) and delocalized regimes, blue and red lines show Poisson and Wigner-Dyson distribution.
    (d)~Comparison of average fractal dimensions $D_2$ for different ensembles depending on diagonal disorder $W$. Dependence of fractal dimension on energy $E$ and symmetry (c)~conserve and (d)~non-conserve diagonal disorder $W$.}
    \label{fig:2}
\end{figure*}

For the superconducting ensemble on the graph with adjacency matrix $A$, the Bogoliubov-de Gennes Hamiltonian has the following form. 
\begin{equation}\label{eq:BdG}
    H_{BdG}= 
    \begin{pmatrix}
        A & B & \Delta^{\ua\ua} & \Delta \\
        B^T & C & -\Delta^T & \Delta^{\da\da} \\
        \Delta^{\ua\ua} & -\Delta^T& -A^T & -B \\
        \Delta & \Delta^{\da\da} & -B^T & -C \\
    \end{pmatrix}
\end{equation}
where $B$ is the matrix for terms with the destruction and creation of electrons with opposite spins $c^{\dagger}_{\alpha,\uparrow}c_{\beta,\downarrow}$ and $c^{\dagger}_{\alpha,\downarrow}c_{\beta,\downarrow,\uparrow}$. The presence of $B$ and $\Delta^{\ua\ua}$, $\Delta^{\da\da}$ corresponds to \cs{D} and \cs{DIII} symmetry classes; for \cs{C} and \cs{CI} classes, they are equal to zero. The class \cs{C} has spin-reversal symmetry, but time is broken, while \cs{CI} conserves both symmetries.
Furthermore, \cs{CI} is the only class that has $\beta=1$ for random matrix representation, as in \cs{AI} and \cs{BDI}. Consequently, further \cs{CI} is studied. The gap matrix is restricted to only having constant elements on the diagonal, $\Delta_{\alpha\beta} = \Delta_0 \delta_{\alpha\beta}$, where $\delta_{\alpha\beta}$ is the Kronecker symbol.
Figure~\ref{fig:2} shows the results of numerical calculations for the superconducting Hamiltonian with Anderson diagonal disorder: 
\begin{multline}\label{eq:BdG+A}
    \hat{H}_{BdG+A} =  \sum_n \varepsilon_{n} 
    (c^{\dagger}_{n\ua}c_{n\ua} + c^{\dagger}_{n\da}c_{n\da}) +
    \sum_{nm} A_{nm}(c^{\dagger}_{n\ua}c_{m\ua}+c^{\dagger}_{n\da}c_{m\da}) + \\ +
    \frac{1}{2}\Delta_0 \sum_{n} c_{n\ua}c_{n\da} +
    \frac{1}{2}\Delta_0^* \sum_{n} c^{\dagger}_{n\ua}c^{\dagger}_{n\da} \ .
\end{multline}
The matrices $A_{nm}$ with size $N \times N$ belong to the RRG ensemble.

Figure~\ref{fig:2}(a) shows the density of states without the diagonal disorder. A large amount of states are concentrated near the gap. Since the studied Hamiltonian has spin-rotation symmetry, for every state, the degree of degeneracy is two. Furthermore, from the characteristic equation, $\det(H -\lambda I) = 0$, where $I$ is the identity matrix and $H$ in the form~(\ref{eq:BdG+A}), the number of independent states is reduced by half.
After the product of block matrices in the Hamiltonian, the characteristic equation has the following form. 
\begin{equation}
    \det{\left(A^2-(\lambda^2 - \Delta_0^2) I \right)^2} = 0 \ .
\end{equation}
Let $\nu = \mu^2 = (\lambda^2 - \Delta_0^2)$, then the eigenvalues satisfy equations $\det( A \pm \mu I) = 0$. Hence, the number of independent parameters is reduced by half, i.e. a quarter from the number of all states. 

For the ensemble of RRG $\mu$, it has the Kesten-McKay distribution. The connection between spectrum eigenvalues gives the connection between distribution for $\nu$ and $\mu$. It is an analog of the $\chi^2$-distribution for one degree of freedom only for KM.
\begin{equation}
    \rho_{\chi^2} (\nu) = 2 \frac{d \mu}{d \nu} \rho_{KM} (\mu) 
                              = \frac{d \sqrt{4(d-1) - \nu }}
                                     {2 \pi \sqrt{\nu}  (d^2 - \nu )} \ .
\end{equation}
 The density of states for the eigenvalues of $H_{BdG+A}$ is BdG Kesten-McKay or $\chi$-Kesten-McKay with a gap
\begin{equation}\label{eq:BdG-KM}
    \rho_{BdG}(\lambda) = \frac{1}{2} \frac{d \nu}{d \lambda} \rho_{\chi^2} (\nu)
                        = \frac{d \lambda \sqrt{4 (d-1) - (\lambda^2 - \Delta_0^2) }}
                               {2 \pi \sqrt{(\lambda^2 - \Delta_0^2 )} (d^2 - (\lambda^2 - \Delta_0^2 ))} \ .
\end{equation}
In Figure~\ref{fig:2}(a)(b) the histogram shows numerical calculations for the eigenvalues of $H$ in RRG, and the solid black line is calculated by Eq.~(\ref{eq:BdG-KM}).

In Figure~\ref{fig:2}(c) black dots show LSD for distinct states at $W = 0$. It differs from the Wigner-Dyson distribution for GOE and RRG. In this regimes the nodes effectively form two unconnected clusters with adjacency matrices $A$ and $-A$ with eigenvalues $\{\mu_i\}$ and $\{-\mu_i\}$. It leads that some nearest states doesn't have repulsing, so $\lim_{s\to0} P(s) \neq 0$.
A similar distribution can be observed for two unconnected clusters, each of them being Erdős-Rényi graph~\cite{Abuelenin2012Effect}. 

From the fractal dimension point of view in Figure~\ref{fig:2}(d)-(f) we compare two kinds of disorder, spin-rotation conserving and nonconserving, with RRG with different system sizes. For the BdG class, the underlying graph has $N=256$ nodes, but due to spin symmetry and particle-hole decomposition, the total size of the Hamiltonian in matrix form is $N=1024$. In the case of spin invariance, the gap is sustained with the transition and remains at infinite disorder; also, it has a smaller average fractal dimension. When opposite spins have different on-site energies, the gap survives in small disorder but soon vanishes. In that range, an increase of the fractal dimension is connected to an increase of degeneracy. Symmetry breaking leads to the behavior of the fractal dimension becoming coincident with RRG.

\section{Discussion}

In the present study, we investigate Anderson localization on graphs that obey chiral and BdG symmetry classes. For chiral BDI class symmetry, it is shown that for chiral RRG LSD does not differ from RRG. The fractal dimension is large when the models have an equal mean node degree. For BdG on RRG with a constant gap, Eq.~\ref{eq:BdG-KM} is shown for the density of states. If diagonal disorder conserves symmetry, the gap is stable. Otherwise, with increasing disorder, the system tends to behave like an RRG. 

For further investigation, it makes sense to compare critical exponents between graphs of different symmetry classes. The question of what real-life many-body systems can have chiral Hilbert space structures remains open. Using the approach of Sect.~\ref{Sec:Super}, it must have two types of particles ($A$, $B$) with an interaction term $H_I = \sum V c^\dagger_A c^\dagger_A c_B c_B + \sum V c^\dagger_B c^\dagger_B c_A c_A$. Then there are two groups of states distinguished by the parity of the difference in the number of particles of kinds divided by $4$, $(n_A - n_B)/4$. Transitions between states in a group are only possible through another part. 
When the disorder is non-zero only on some nodes, robust disordered states appear in the middle of the spectrum~\cite{Kochergin2024Robust}. The study of their behavior on graphs with symmetry other than orthogonal is interesting.

From the construction of the BdG Hamiltonian adjacency matrix $A$, it takes on a block diagonal form. For each $\ket{\alpha_\da}$, there is a full set of possible states $\ket{\alpha_\ua}$. Since $A$ has transition elements between states with the same $\ket{\alpha_\da}$, it forms clusters or blocks. In Sec.~\ref{Sec:Super}, we use the mean-field approximation for quadratic terms when a pair of electrons has opposite spins. If we account for such quadratic terms, there will be an interaction between the clusters in $A$. Similar graph models appear in exponential networks with a chemical potential for short cycles, where, with an increase in the potential, the random graph can transition to a clustered phase~\cite{Avetisov2016Eigenvalue,Kochergin2023Anatomy}. 

\textbf{Acknowledgements.} We are grateful to Igor Burmistrov for the inspiring question.

\bibliographystyle{hieeetr}
\bibliography{bib.bib}

\end{document}